\pdfoutput=1

\documentclass[11pt]{article}

\usepackage[final]{acl}

\usepackage{times}
\usepackage{latexsym}

\usepackage[T1]{fontenc}

\usepackage[utf8]{inputenc}

\usepackage{microtype}
\usepackage{enumitem}

\usepackage{inconsolata}

\usepackage{graphicx}

\usepackage{adjustbox}
\usepackage{tabularx}
\usepackage{multirow}
\usepackage{booktabs}
\usepackage{hyperref}
\usepackage{tcolorbox}
\usepackage{acl}
\usepackage{amssymb}
\usepackage{amsmath}
\usepackage{subcaption}

\title{Low-Rank and Sparse Model Merging for Multi-Lingual\\ Speech Recognition and Translation}

\author{
\textbf{Qiuming Zhao\textsuperscript{1} \qquad Guangzhi Sun\textsuperscript{2}} \qquad \textbf{Chao Zhang\textsuperscript{1}\thanks{Correspondence}} \\
$^1$Tsinghua University, China \\
$^2$University of Cambridge, United Kingdom \\
\texttt{zqm23@mails.tsinghua.edu.cn, gs534@cam.ac.uk, cz277@tsinghua.edu.cn}
}

\begin{document}
\maketitle
\begin{abstract}
Language diversity presents a significant challenge in speech-to-text (S2T) tasks, such as automatic speech recognition and translation. Traditional multi-lingual multi-task training approaches aim to address this by jointly optimising multiple speech recognition and translation tasks across various languages. While models like Whisper, built on these strategies, demonstrate strong performance, they still face issues of high computational cost, language interference, suboptimal training configurations, and limited extensibility. To overcome these challenges, we introduce LoRS-Merging (low-rank and sparse model merging), a novel technique designed to efficiently integrate models trained on different languages or tasks while preserving performance and reducing computational overhead. LoRS-Merging combines low-rank and sparse pruning to retain essential structures while eliminating redundant parameters, mitigating language interference, and enhancing extensibility. Experimental results across 10 languages demonstrate that LoRS-Merging significantly outperforms multi-lingual multi-task training, sequential training, and other merging methods, achieving over 20\% improvement in normalised performance. Our findings suggest that model merging, particularly LoRS-Merging, is a scalable and effective complement to traditional multi-lingual training strategies for S2T applications\footnote{The detailed data and code will be released at [URL]}.

\end{abstract}

\section{Introduction}
Language diversity poses a significant challenge in speech-to-text (S2T) tasks, such as automatic speech recognition (ASR) \cite{prabhavalkar2023end} and speech translation (ST) \cite{xu2023recent}. 
With over 7,000 languages spoken worldwide, developing robust S2T systems that generalise across varied linguistic structures remains a fundamental research goal \cite{liu2024recent,cheng2023mu2slam,sun2023towards,saif2024m2asr,wang2021voxpopuli,le2021lightweight}.
The transition from pipeline systems to end-to-end (E2E) models \cite{chan2016listen,gulati2020conformer,barrault2023seamlessm4t} has marked a paradigm shift in S2T tasks, enabling direct speech-to-text mapping across multiple languages within a unified framework.
A prominent example is Whisper \cite{radford2023robust}, an advanced multi-lingual speech model trained on a large-scale, diverse dataset covering multiple languages and tasks.
Despite these advances, existing multi-lingual models still encounter significant challenges in scalability, efficiency, and performance trade-offs.

To address these challenges, multi-lingual training strategies \cite{saif2024m2asr,xiao2021adversarial,bai2018source} have been adopted, aiming to enhance model generalisation across languages.
These approaches typically rely on joint optimisation of diverse S2T tasks across multiple languages, leveraging shared representations to improve performance.
Nevertheless, multi-lingual training is subject to inherent limitations, including substantial training costs, complex model configurations, and limited access to training data across multiple languages and tasks.
Moreover, when handling new languages, the training methods typically require training from scratch, leading to poor extensibility.

To mitigate these issues, this paper proposes to use model merging \cite{ilharcoediting,yang2024model,khan2024sok} to integrate models trained on different languages or tasks while maintaining performance and reducing computational overhead.
Model merging merges the parameters of multiple separate models with different capabilities to build a universal model.
With its high flexibility, model merging enables the seamless incorporation of new languages or tasks without the need for retraining the entire model.
Additionally, since model merging allows models for different languages or tasks to be trained independently, it can effectively alleviate negative transfer issues \cite{wang2019characterizing,zhang2022survey,wang2020negative} commonly observed in multi-lingual training.
This training independence also enables optimal training configurations for each language or task to improve performance, instead of the unified settings required in multi-lingual training.

Moreover, we propose \textbf{Lo}w-\textbf{R}ank and \textbf{S}parse model \textbf{Merging} (LoRS-Merging), which uses a low-rank component to capture the compact structure and a sparse component to capture the scattered details in the weights.
LoRS-Merging retains effective parts of structure and details while reducing redundant parts to reduce language interference.
Specifically, coarse-grained singular value pruning is used to retain the low-rank structure, while fine-grained magnitude pruning is used to remove redundant details.
The main contribution of this paper can be summarised as follows.
\begin{itemize}[itemsep=-1pt, leftmargin=*]
\item To the best of our knowledge, this work is the first to explore model merging for speech-to-text models. Specifically, we treat speech tasks (recognition and translation) and different languages as two separate merging levels and explore different hierarchies for model merging.
\item We propose LoRS-Merging, which exploits the low-rank structure and sparsity of model weights to minimise model redundancy and language conflicts as well as providing an efficient way to incorporate new knowledge from a task- or language-specialised model.
\item Experiments are performed across 10 languages, where LoRS-Merging significantly outperforms multi-lingual multi-task training, sequential training, and other merging methods, achieving over 20\% improvement in normalised performance.

\end{itemize}

\section{Related Work}

\subsection{Multi-Lingual ASR and ST}
Multi-lingual speech models inherently face a trade-off between knowledge sharing and negative interference.
Early studies adopted hand-picked sub-network sharing strategies, such as language-specific decoders \cite{dong2015multi}, attention heads \cite{zhu2020multilingual}, and layer norm/linear transformation \cite{zhang2020improving}.
Recent research has shifted toward approaches such as mixture-of-experts \cite{kwon2023mole,wang2023language}, adapters \cite{le2021lightweight,kannan2019large}, and pruning \cite{lu2022language,lai2021parp}.
To enhance multi-lingual representation learning, language tokens \cite{johnson2017google}, embeddings \cite{di2019one} or output factorisations \cite{zhang2023umluniversalmonolingualoutput} are introduced to encode language identity, helping the model distinguish between languages.

The more effective approach is to adopt multi-lingual training strategies, such as multi-objective optimisation \cite{saif2024m2asr,zhang2022streamingendtoendmultilingualspeech}, adversarial learning \cite{xiao2021adversarial}, meta learning \cite{hsu2020meta}, and reinforcement learning \cite{bai2018source}.
Moreover, large-scale pretraining by leveraging massive amounts of multi-lingual and multi-task data enables models to learn robust and transferable representations across languages, e.g. Whisper \cite{radford2023robust}, SeamlessM4T \cite{barrault2023seamlessm4t}, and AudioPaLM \cite{rubenstein2023audiopalm}. LoRS-Merging, as an efficient post-training method proposed in this paper, further advances multi-lingual ASR and ST based on pretrained speech models.

\subsection{Model Merging}
Model merging \cite{yang2024model,khan2024sok} is an efficient post-training technique that integrates knowledge from models trained on different domains.
One stream of research focuses on the loss landscape geometry \cite{khan2024sok} and studies the linear mode connectivity (LMC) \cite{frankle2020linear,draxler2018essentially} property that demonstrates the existence of a linearly connected path between local minima within the same loss basin.
Many studies \cite{nagarajan2019uniform,izmailov2018averaging,frankle2020linear} indicate that if two neural networks share part of their optimisation trajectory, such as different finetuned models from the same pretrained model, they typically satisfy LMC, allowing interpolation without sacrificing accuracy and forming the basis of our model merging method. 
For local minima in different loss basins, inspired by the permutation invariance \cite{entezarirole} of neural networks, neuron alignment techniques \cite{ainsworthgit,singh2020model,tatro2020optimizing} can be used to place them into the same basin, thereby reducing merging loss.

Another stream considers the model spaces, including activation spaces and weight spaces.
Research on activation spaces seeks to align the output representations or loss of the merged model with those of each single model as closely as possible \cite{yangrepresentation,wei2025modeling,xiong2024multi}. Studies based on weight spaces aim to localise effective parameters or remove redundant parameters to resolve task interference.
TALL-masks \cite{wanglocalizing} and Localise-and-Stitch \cite{he2024localize} optimise binary masks to localise sparse and effective task-specific parameters.
TIES-Merging \cite{yadav2024ties} and DARE \cite{yu2024language} perform magnitude or random pruning on each single model to reduce redundancy at the detailed parameter level.
TSV-M \cite{gargiulo2024task}, on the other hand, adopts singular value pruning to reduce redundancy at the structural level.
In contrast, LoRS-Merging explores weight space merging by considering not only the detailed parameter redundancy as well as maintaining the effective structure of the weight space.

\section{Methodology}
\begin{figure*}[t]
  \vspace{-1cm}
  \includegraphics[width=\linewidth]{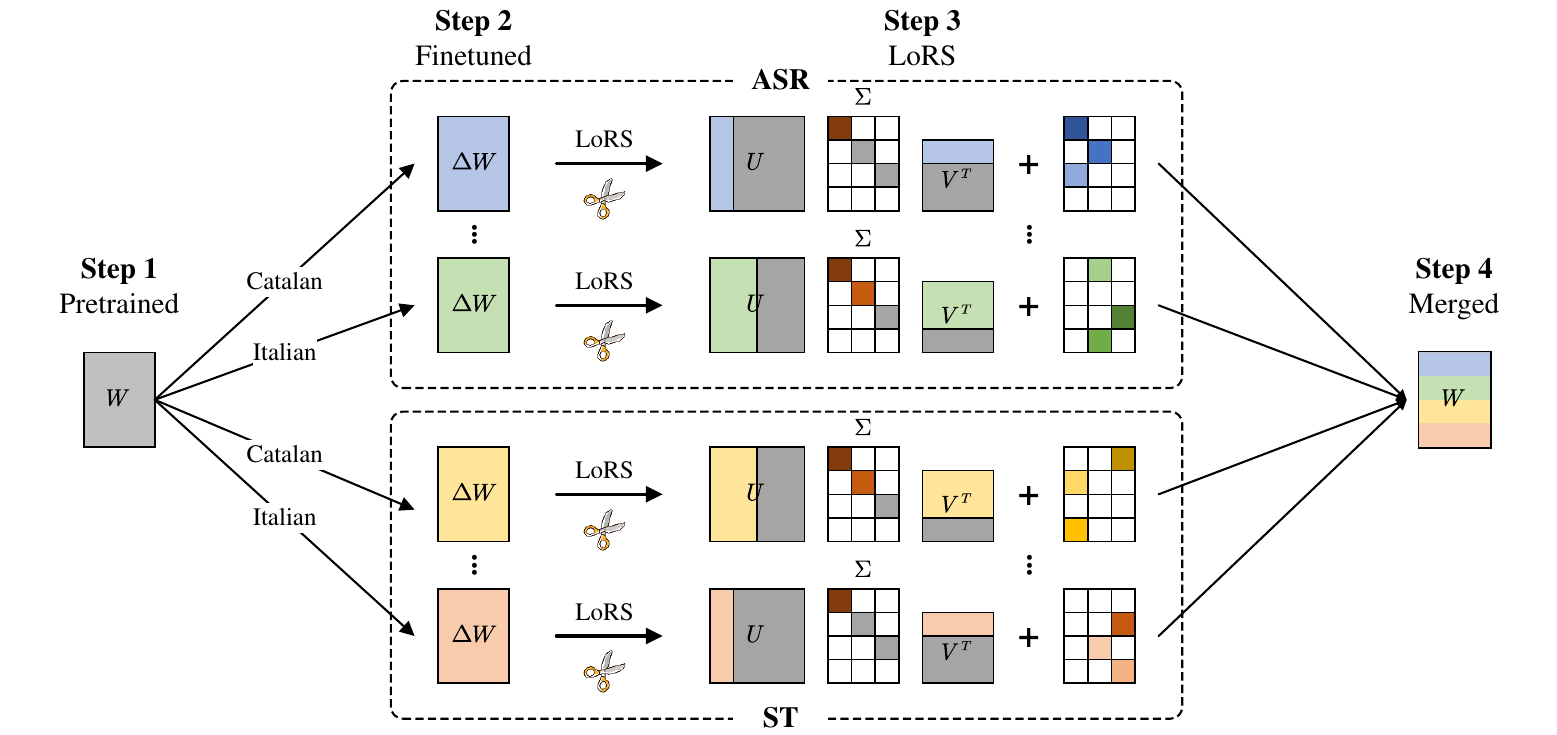}
  \caption{Model merging process with the proposed LoRS-Merging for speech models on multi-lingual ASR and ST tasks.
  In step 1, a suitable pretrained speech model is selected.
  In step 2, the pretrained model is finetuned with the task-language-specific data.
  In step 3, apply LoRS to the delta parameters to reduce model redundancy.
  In step 4, merge the delta parameters to get a multi-lingual and multi-task merged model.}
  \vspace{-0.5cm}
  \label{fig:overview}
\end{figure*}

\subsection{Preliminaries}
\subsubsection{Task Arithmetic}
Among diverse model merging methods, Task Arithmetic (TA) \cite{ilharcoediting} has become a fundamental technique in this field due to its simplicity and effectiveness.
TA introduces the concept of "task vector", defined as the delta parameter derived by subtracting pretrained weights from finetuned weights.
By performing simple arithmetic operations on task vectors, TA enables task learning, forgetting, and analogising.

Assume that $\theta = \{W_l\}_{l=1}^{L}$ represents the parameters of the model, where $W_l$ is the weight of $l$-th layer, and $L$ is the total number of layers.
Given a pretrained model $\theta_0$ and a model $\theta_i$ finetuned on task $t_i$, the task vector is computed as $\tau_i = \theta_i - \theta_0$.
Multiple task vectors can be summed to form a multi-task model, expressed as $\theta_\text{merged} = \theta_0 + \lambda \sum_{i=1}^{n} \tau_i$,  where $\lambda$ is a scaling coefficient for the task vectors.

\subsubsection{Pruning}
\label{sec:pruning}
Given that neural networks are typically over-parameterised and exhibit high redundancy, a considerable number of neurons or connections can be pruned without affecting accuracy \cite{lecun1989optimal}.
In model merging, pruning methods can reduce redundant parameters to mitigate task interference, thereby improving the merging performance.

\textbf{Magnitude Pruning} (MP) is an unstructured pruning method that prunes connections based on the magnitude of parameters as a measure of importance.
Specifically, MP prunes the parameters according to a specific ratio $p$, as follows.
\begin{equation}
M_{ij} = 
\begin{cases} 
1 & \text{if } |w_{ij}| \in \text{top } p\% \\
0 & \text{o.w.}
\end{cases}
\end{equation}
\begin{equation}
W_\text{pruned} = M \odot W
\end{equation}
where $W, M \in \mathbb{R}^{d \times k}$, and $\odot$ denotes the element-wise multiplication.
However, MP only focuses on the redundancy at the parameter level, overlooking the crucial structural information, which may lead to the disruption of the weight structure.

\textbf{Singular Value Pruning} (SVP) is a structured pruning method that removes smaller singular values and their corresponding singular vectors.
In particular, SVP retains only the top $r$ singular values while discarding the others.
\begin{equation}
W = U \Sigma V^T
\end{equation}
\begin{equation}
W_\text{pruned} = U_r \Sigma_r V_r^T
\end{equation}
where $U \in \mathbb{R}^{d \times d}$ and $V \in \mathbb{R}^{k \times k}$ are the left and right singular vector matrices of $W$, and $U_r$, $V_r$ denote their first $r$ columns.
Although SVP preserves a compact weight structure, its coarse pruning granularity makes it challenging to reduce redundancy at a fine-grained parameter level.

\subsection{Model Merging for Speech Models}

The model merging process for speech model on S2T tasks with LoRS-Merging as an example is shown in Fig. \ref{fig:overview}, which comprises four steps. In step 1, a suitable pretrained speech model is selected. In step 2, for each target language and target task combination, e.g. Catalan ASR, the pretrained model is finetuned with the task-language-specific data and the delta weight is obtained. In step 3, weight pruning is applied to remove redundant and conflicting delta parameters. In step 4, task arithmetic is applied to combine pruned delta weights into each single merged matrix and hence obtain the merged model.

Model merging allows new language or task knowledge to be integrated into the model in a flexible post-training manner.
Compared to multi-lingual or multi-task training methods, model merging is a simpler and more efficient approach, enabling the seamless incorporation of new languages or tasks without the need for retraining. Additionally, due to its training independence, it mitigates language conflicts and provides optimal training configurations for each language or task to improve performance.
Compared to sequential training, model merging eliminates the need for additional training data to avoid catastrophic forgetting.
Our experiments thoroughly demonstrate these benefits.

\subsection{Low-Rank and Sparse Model Merging}
\begin{figure}[t]
  \includegraphics[width=\linewidth]{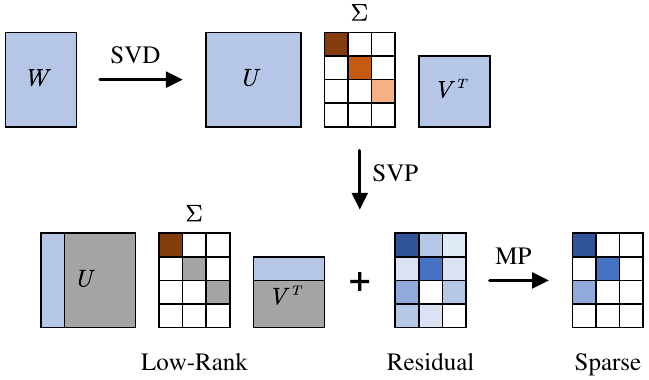}
  \caption{Illustration of LoRS-Merging method in detail. SVD stands for singular value decomposition and SVP for singular value pruning. MP is magnitude pruning operating on residual of the original weight matrix and the low-rank matrix.}
  \label{fig:lorsmerge}
  \vspace{-0.4cm}
\end{figure}

The weights of neural networks contain information on both structure and details. 
Structural information is coherent, compact, and coarse-grained, whereas detail information is incoherent, scattered, and fine-grained.
Both structural and detail information include effective and redundant parts.
To reduce redundant parts in both the structure and detail aspects of the weights while retaining effective parts, the LoRS-Merging method is introduced as shown in detail in Fig. \ref{fig:lorsmerge}, which exploits the combination of low-rank structure by SVP and sparsity by MP.
SVP performs coarse-grained pruning at the structure level, while MP enables fine-grained pruning at the detail level.

In the implementation, we approximate the original weights as the sum of a low-rank component and a sparse component, where the low-rank component captures the compact structure, and the sparse component captures the scattered details, as shown in Eqn. \eqref{eq:sum}.
\begin{equation}
W \approx L + S
\label{eq:sum}
\end{equation}
where $L$ represents the low-rank component, and $S$ represents the sparse component.
Specifically, \( L \) is the low-rank matrix obtained by retaining the top \( r \) singular values and their corresponding singular vectors from \( W \):
\begin{equation}
L = U_r \Sigma_r V_r^T
\end{equation}
and $S$ is the sparse matrix obtained by performing MP on the residual of $W$ and $L$:
\begin{equation}
S = M \odot (W - L)
\end{equation}
To simplify the description, we refer to this entire process as $\text{LoRS}(\cdot)$. In this manner, SVP decouples the structure and details of the weight, preserving a compact structure while allowing fine-grained MP to remove redundant parts in the details.

For each model finetuned on single specific language or task data, we apply $\text{LoRS}(\cdot)$ to its task vector as a preprocessing step to reduce language or task interference in model merging.
A multi-lingual or multi-task model can be achieved through simple merging, expressed as: 
\begin{equation}
\theta_\text{merged} = \theta_0 + \lambda \sum_{i=1}^{n} \text{LoRS}(\tau_i)
\end{equation}

\section{Experimental Setup}
\subsection{Data}
\textbf{CoVoST-2} \cite{wang2020covost} is a large-scale multi-lingual ST corpus based on Common Voice.
It covers translations from English into 15 languages and from 21 languages into English, with a total of 2,880 hours of speech from 78k speakers.
We selected 5 high-resource languages and 5 low-resource languages as two language sets to investigate their ASR tasks and the from X to English ST tasks.
The high-resource language set includes Catalan (ca), German (de), Spanish (es), French (fr), and Italian (it), while the low-resource language set includes Indonesian (id), Dutch (nl), Portuguese (pt), Russian (ru), and Swedish (sv).
Due to the more abundant data in the high-resource language set, our main experimental results are obtained on the high-resource language set, while the low-resource language set serves as an auxiliary evaluation set.
To balance the amount of data across different languages, we fixed the duration of traning data for each language, with 5 hours for the high-resource language set and 1 hour for the low-resource language set.
The dev and test sets of both language sets are 1 hour in duration.

\subsection{Model and Training Specifications}
{\bf Whisper} \cite{radford2023robust} is a general-purpose multi-lingual ASR and ST model, a Transformer-based model trained on 680k hours of diverse audio.
We chose the small version as the foundation model for the experiments because it achieves a good balance between performance and cost.
It has 244 million parameters, with the encoder and decoder each consisting of 12 Transformer blocks.
The weight matrices of the attention layers are all 768 by 768, and the MLP layers are 768 by 3072.

For each language-specific or task-specific finetuned model, we use a different, optimal learning rate for each during training, and these models are subsequently used for model merging.
Finetuning involves updating all parameters.
We choose Adam as the optimiser, set the batch size to 8, the accumulation iterations to 4, and train for 10 epochs.
The beam size for decoding is set to 20 across all languages and tasks.
We use Sclite and SacreBLEU tools to score the ASR and ST results, respectively.
In addition, we perform statistical significance testing using a paired bootstrap test with 1,000 resampling iterations, each sampling 300 examples with replacement, and report the results in the caption of each table.
See Appendix \ref{sec:appendix_setup} for more details on the experimental setup. Our experiments are performed on a single RTX 4090 GPU where training on one language and one task with 5 hours of speech data requires 1 hour.

\subsection{Baseline and Merging Methods}
We use the \textbf{pretrained model} as the baseline and \textbf{multi-lingual multi-task training} as the stronger baseline, which is trained on data mixed from both multi-lingual and multi-task sets.
Note that finetuned models are typically available, so model merging requires no finetuning and only adjusts merging coefficients on a small development set. Even if finetuning is needed, the complex training configurations of multi-lingual and multi-task training require more hyperparameter tuning steps. Overall, model merging consumes significantly fewer computational resources than multi-lingual and multi-task training.

In addition to LoRS-Merging, we investigate the following model merging methods:

\textbf{Weight Averaging} (WA) \cite{wortsman2022model} merges multiple single models by a uniform averaging of their weights.

\textbf{Task Arithmetic} (TA) \cite{ilharcoediting} uses a scaling factor, estimated on a small development set, to weight multiple task vectors.



\textbf{TIES-Merging} \cite{yadav2024ties} merges single models via Trim, Elect, and Disjoint Merge steps to reduce parameter interference.

\textbf{DARE} \cite{yu2024language} drops and rescales delta parameters to mitigate parameter interference.

\textbf{TSV-M} \cite{gargiulo2024task} proposes Task Singular Vectors and reduces structural redundancy to reduce task interference.

In addition, we report the normalised performance difference defined in Eqn. \eqref{eq:norm}.
\begin{equation}
    \Delta_\text{norm} = \frac{|M - M_\text{pretrained}|}{|M_\text{finetuned} - M_\text{pretrained}|} \times 100\%
    \label{eq:norm}
\end{equation}
where $M$ is the performance metric of the target system, $M_\text{pretrained}$ and $M_\text{finetuned}$ are for the pretrained (baseline performance) and finetuned (topline performance) systems respectively. \emph{Note that $\Delta_\text{norm}$ better reflects the performance gains for model merging since the pretrained system already achieves competitive performance.}

\section{Evaluation Results and Analysis}

\subsection{Multi-Lingual Model Merging}

\begin{table}[t]
  \caption{Multi-lingual ASR model merging. Avg. denotes average WER. $\ast$ LoRS-Merging outperforms all others in $\Delta_{\text{norm}}$ by >20\% ($p<0.05$).}
  \label{tab:multi-lingual-ASR}
  \centering
  \begin{adjustbox}{width=\columnwidth}
  \begin{tabular}{l|ccccc|cc}
    \toprule
    \multicolumn{1}{l}{\multirow{2}{*}{\textbf{System}}} & \multicolumn{7}{c}{\textbf{WER$\downarrow$}} \\
     & ca & de & es & fr & it & Avg. & $\Delta_{\text{norm}}$ \\
    \midrule
    Pretrained & 20.6 & 19.6 & 14.7 & 24.5 & 19.4 & 19.88 & - \\
    Finetuned & 19.5 & 19.7 & 14.4 & 22.1 & 19.2 & 19.05 & 100.0\% \\
    \midrule
    Multi-lingual training & 17.1 & 21.8 & 15.1 & 22.6 & 21.9 & 19.69 & 22.9\% \\
    Sequential training & 20.6 & 19.6 & 14.6 & 24.4 & 19.4 & 19.84 & 4.8\% \\
    \midrule
    Weight Averaging & 19.1 & 19.1 & 14.2 & 24.5 & 20.3 & 19.55 & 39.8\% \\
    Task Arithmetic & 19.1 & 18.8 & 13.9 & 24.0 & 19.8 & 19.23 & 78.3\% \\
    TIES-Merging & 19.3 & 19.3 & 13.9 & 23.8 & 18.1 & 18.99 & 107.2\% \\
    DARE & 18.9 & 18.9 & 13.9 & 23.8 & 19.8 & 19.16 & 86.7\% \\
    TSV-M & 19.5 & 19.5 & 14.1 & 23.5 & 18.4 & 19.10 & 94.0\% \\
    LoRS-Merging & 18.9 & 18.8 & 13.9 & 23.6 & 18.1 & \textbf{18.77} & \textbf{133.7\%} \\
    \bottomrule
  \end{tabular}
  \end{adjustbox}
\end{table}

\begin{table}[t]
  \caption{Multi-lingual ST model merging. Avg. denotes average BLEU. $\ast$ LoRS-Merging outperforms all others in $\Delta_{\text{norm}}$ by >20\% ($p<0.05$).}
  \label{tab:multi-lingual-ST}
  \centering
  \begin{adjustbox}{width=\columnwidth}
  \begin{tabular}{l|ccccc|cc}
    \toprule
    \multicolumn{1}{l}{\multirow{2}{*}{\textbf{System}}} & \multicolumn{6}{c}{\textbf{BLEU$\uparrow$}} \\
     & ca & de & es & fr & it & Avg. & $\Delta_{\text{norm}}$ \\
    \midrule
    Pretrained & 21.1 & 24.1 & 28.6 & 26.8 & 26.8 & 25.48 & - \\
    Finetuned & 22.6 & 24.6 & 29.2 & 27.2 & 27.3 & 26.18 & 100.0\% \\
    \midrule
    Multi-lingual training & 21.4 & 24.4 & 28.8 & 26.8 & 27.2 & 25.72 & 34.3\% \\
    Sequential training & 21.5 & 24.3 & 28.9 & 26.9 & 27.3 & 25.78 & 42.9\% \\
    \midrule
    Weight Averaging & 22.3 & 24.1 & 28.6 & 27.2 & 26.9 & 25.82 & 48.6\% \\
    Task Arithmetic & 22.1 & 24.3 & 28.9 & 27.3 & 26.8 & 25.88 & 57.1\% \\
    TIES-Merging & 22.1 & 24.7 & 29.0 & 27.1 & 26.9 & 25.96 & 68.6\% \\
    DARE & 22.1 & 24.5 & 28.9 & 27.2 & 26.8 & 25.90 & 60.0\% \\
    TSV-M & 22.0 & 24.6 & 29.0 & 27.3 & 26.8 & 25.94 & 65.7\% \\
    LoRS-Merging & 22.4 & 24.8 & 28.9 & 27.6 & 27.0 & \textbf{26.14} & \textbf{94.3\%} \\
    \bottomrule
  \end{tabular}
  \end{adjustbox}
  \vspace{-0.3cm}
\end{table}

First, we investigate the merging of finetuned models for different languages on the same task, which corresponds to \textit{multi-lingual single-task} learning.

\textbf{Language knowledge interference yields imbalanced improvements}: Table \ref{tab:multi-lingual-ASR} shows the multi-lingual results of the ASR task with the high-resource language set.
On average, multi-lingual training slightly improves the pretrained model but significantly underperforms the finetuned models and merging methods.
This may be due to negative interference between the knowledge of different languages, leading to gradient conflicts during training \cite{wang2020negative}. From a per-language perspective, it is observed that ca and fr achieve the largest improvements during finetuning while still showing significant improvements in multi-lingual training, whereas languages with smaller improvements during finetuning exhibit a substantial performance drop in multi-lingual training, even worse than the pretrained model. This indicates a strong language conflict in multi-lingual training, with ca and fr dominating. Additionally, we observe that the optimal learning rates for finetuned models vary significantly across languages (see Appendix \ref{sec:appendix_setup}), while the unified learning rate configuration required by multi-lingual training prevents each language from reaching its optimal performance.
Moreover, the substantially inferior performance of sequential training indicates the presence of catastrophic forgetting.

\textbf{Model merging mitigates language conflicts}: 
In contrast, model merging methods show significant improvements across almost all languages, demonstrating reduced conflict and better stability. 
Among model merging methods, TA outperforms WA due to its flexible scaling factor.
TIES-Merging and DARE reduce redundancy at the detail level, while TSV-M addresses redundancy at the structural level, thereby achieving obvious improvements over TA.
Furthermore, LoRS-Merging reduces redundancy at the detail level while preserving critical structures, resulting in the best performance.

Table \ref{tab:multi-lingual-ST} provides the multi-lingual results on ST task with the high-resource language set.
The main conclusion is consistent with the ASR task: model merging methods still significantly outperform multi-lingual training and sequential training, with LoRS-Merging achieving the best performance, demonstrating superior multi-lingual and continual learning capabilities.

\subsection{Multi-Task Model Merging}
\begin{table}[t]
    \centering
    \setlength{\tabcolsep}{2pt}
    \caption{Multi-task model merging performed on each language independently. Avg. denotes average WER/BLEU. $\ast$ LoRS-Merging outperforms all others in $\Delta_{\text{norm}}$ by >20\% ($p<0.05$). Per-language results are shown in Appendix \ref{sec:appendix_detail}.}
    \begin{adjustbox}{width=\columnwidth}
    \begin{tabular}{l|cc|cc}
    \toprule
    \textbf{System} & \textbf{Avg. WER}$\downarrow$ & \textbf{$\Delta_{\text{norm}}$} & \textbf{Avg. BLEU}$\uparrow$ & \textbf{$\Delta_{\text{norm}}$} \\
    \midrule
    Pretrained & 19.88 & - & 25.48 & - \\
    Finetuned & 19.05 & 100.0\% & 26.18 & 100.0\% \\
    \midrule
    Multi-task training & 19.00 & 106.0\% & 25.90 & 60.0\% \\
    Sequential training & 18.95 & 112.0\% & 26.12 & 91.4\% \\
    \midrule
    Weight Averaging & 18.84 & 125.3\% & 26.18 & 100.0\% \\
    Task Arithmetic & 18.76 & 134.9\% & 26.30 & 117.1\% \\
    TIES-Merging & 18.60 & 154.2\% & 26.38 & 128.6\% \\
    DARE & 18.71 & 141.0\% & 26.28 & 114.3\% \\
    TSV-M & 18.70 & 142.2\% & 26.40 & 131.4\% \\
    LoRS-Merging & \textbf{18.39} & \textbf{179.5\%} & \textbf{26.56} & \textbf{154.3\%} \\
    \bottomrule
    \end{tabular}
    \end{adjustbox}
    \label{tab:multi-task}
    \vspace{-0.3cm}
\end{table}

\begin{table*}[t]
  \vspace*{-1cm} 
  \caption{Multi-lingual multi-task model merging. Avg. denotes average WER/BLEU.}
  \label{tab:multi-lingual multi-task}
  \centering
  \begin{adjustbox}{width=0.95\textwidth}
  \begin{tabular}{l|ccccc|cc|ccccc|cc}
    \toprule
    \multicolumn{1}{l}{\multirow{2}{*}{\textbf{System}}} & \multicolumn{6}{c}{\textbf{WER$\downarrow$}} & \multicolumn{6}{c}{\textbf{BLEU$\uparrow$}} \\
     & ca & de & es & fr & it & Avg. & $\Delta_{\text{norm}}$ & ca & de & es & fr & it & Avg. & $\Delta_{\text{norm}}$ \\
    \midrule
    Pretrained & 20.6 & 19.6 & 14.7 & 24.5 & 19.4 & 19.88 & - & 21.1 & 24.1 & 28.6 & 26.8 & 26.8 & 25.48 & - \\
    Finetuned & 19.5 & 19.7 & 14.4 & 22.1 & 19.2 & 19.05 & 100.0\% & 22.6 & 24.6 & 29.2 & 27.2 & 27.3 & 26.18 & 100.0\% \\
    \midrule
    ML and MT training & 20.5 & 19.7 & 14.6 & 24.5 & 19.4 & 19.86 & 2.4\% & 21.3 & 24.3 & 28.3 & 27.1 & 26.9 & 25.58 & 14.3\% \\
    \midrule
    ML and MT Task Arithmetic & 18.9 & 19.2 & 14.1 & 23.7 & 18.4 & 18.96 & 110.8\% & 22.2 & 24.4 & 29.0 & 27.3 & 26.9 & 25.96 & 68.6\% \\
    ML and MT LoRS-Merging & 18.7 & 19.1 & 14.0 & 23.8 & 18.0 & 18.82 & 127.7\% & 22.2 & 24.8 & 29.0 & 27.5 & 27.0 & 26.10 & 88.6\% \\
    \midrule
    MT training & 17.0 & 19.7 & 14.4 & 24.2 & 19.4 & 19.00 & - & 22.3 & 24.6 & 28.7 & 27.0 & 26.9 & 25.90 & - \\
    $\hookrightarrow$ + ML Task Arithmetic & 18.1 & 19.0 & 14.2 & 24.5 & 20.6 & 19.37 & 61.4\% & 22.7 & 24.7 & 28.6 & 27.3 & 26.5 & 25.96 & 68.6\% \\
    $\hookrightarrow$ + ML LoRS-Merging & 18.1 & 19.0 & 14.1 & 24.2 & 20.3 & 19.23 & 78.3\% & 22.4 & 24.5 & 29.1 & 27.6 & 26.7 & 26.06 & 82.9\% \\
    \midrule
    ML training & 17.1 & 21.8 & 15.1 & 22.6 & 21.9 & 19.69 & - & 21.4 & 24.4 & 28.8 & 26.8 & 27.2 & 25.72 & - \\
    $\hookrightarrow$ + MT Task Arithmetic & 17.1 & 18.5 & 13.3 & 22.7 & 18.0 & 18.00 & 226.5\% & 22.6 & 25.0 & 29.2 & 27.5 & 26.9 & 26.24 & 108.6\% \\
    $\hookrightarrow$ + MT LoRS-Merging & 16.9 & 18.3 & 13.3 & 22.4 & 17.8 & \textbf{17.82} & \textbf{248.2\%} & 22.8 & 25.2 & 29.3 & 27.6 & 27.0 & \textbf{26.38} & \textbf{128.6\%} \\
    \bottomrule
  \end{tabular}
  \end{adjustbox}
  \vspace{-0.3cm}
\end{table*}

Next, we merge finetuned models for different tasks (ASR and ST) with the same language which corresponds to \textit{multi-task single-language} learning.

\textbf{ASR and ST tasks for the same language can mutually benefit from each other}: Table \ref{tab:multi-task} presents the multi-task results with the high-resource language set.
In general, multi-task training performs similarly to finetuned models on ASR but is a lot worse on ST. This is likely due to the substantial differences in optimal hyper-parameter configurations between the two tasks.
Sequential training performs similarly to finetuned models overall, as it also benefits from training independence.
Model merging methods clearly outperform finetuned models, which not only demonstrates their effectiveness but also shows the mutual benefits between ASR and ST.
In terms of performance gains, the improvement in ASR is greater than in ST. We attribute this to the fact that ASR is inherently simpler than ST and can be viewed as a step in the ST task.
Furthermore, as before, model merging methods combined with pruning further improve performance, and the proposed LoRS-Merging achieves the best performance across the table.

\subsection{Multi-Lingual Multi-Task Model Merging}

Then, we investigate the merging of finetuned models for both different languages and tasks, which correspond to \textit{multi-lingual (ML) and multi-task (MT)} learning.
Specifically, we explore 4 different training and merging settings: 

\textbf{ML and MT training}: Finetuning on all languages and both tasks jointly.

\textbf{ML and MT merging}: Finetuning on each language for each task separately and merging all.

\textbf{MT training and ML merging}: Finetuning both tasks jointly for each language, and merging models from different languages.

\textbf{ML training and MT merging}: Finetuning on all languages jointly for each task, and merging models from different tasks.

Table \ref{tab:multi-lingual multi-task} displays the multi-lingual and multi-task results with the high-resource language set.
Multi-lingual and multi-task training shows little improvement over the pretrained model, due to language interference during training and the use of a unified training configuration for all languages and tasks.
Nevertheless, the performance of multi-lingual and multi-task merging is on par with that of finetuned models, further underscoring the superiority of model merging.
ML training followed by MT merging achieves the best performance, even significantly outperforming finetuned models.
Although we did not observe the same phenomenon on the low-resource language set, this suggests the potential of using a combination of training and merging to achieve better performance.
We provide additional experiments on the low-resource language set in Appendix \ref{sec:appendix_low} to demonstrate the robustness and generalisability of model merging and LoRS-Merging.

\subsection{Effect of Numbers of Languages}

\begin{figure}[t]
    \centering
    \includegraphics[width=1.0\linewidth]{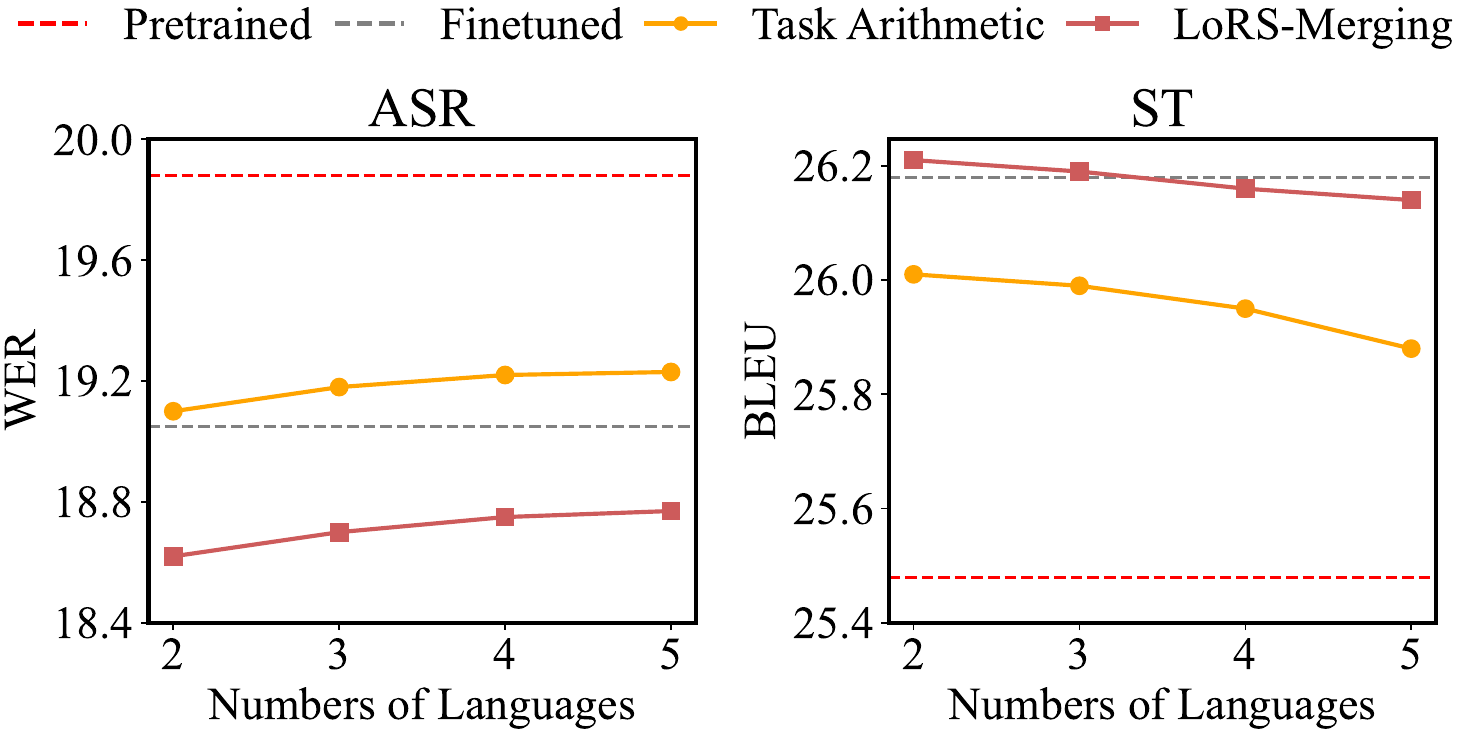}
    \caption{WER and BLEU against the number of languages. Performance is averaged across all languages and all training runs of language combinations.}
    \label{fig:language-number}
    \vspace{-0.3cm}
\end{figure}

To further demonstrate the robustness of LoRS-Merging to language selection, experiments are performed using different numbers of languages. Figure \ref{fig:language-number} shows the average performance across all languages and all training runs with possible combinations of 2, 3, 4 or 5 languages.

\textbf{LoRS-Merging improvements are consistent across different numbers of languages}: As the number of languages increases, the performance of both TA and LoRS-Merging degrades due to negative interference between languages.
LoRS-Merging consistently outperforms TA in both ASR and ST tasks. Notably, in the ASR task, it even surpasses the finetuned models.
This is primarily because finetuned models contain substantial redundancy (see Fig. \ref{fig:svp_mp_asr}), whereas LoRS-Merging reduces redundancy through pruning, leading to significant performance improvements.
Additionally, we observe that the optimal learning rate for the finetuned ASR model is significantly larger compared to the ST task.
This may lead to overfitting in ASR. In contrast, LoRS-Merging improves generalisation through model merging, thus outperforming the finetuned models for the ASR task.

\subsection{Effect of Language Data Scale}

\begin{figure}[t]
  \vspace{-0.5cm}
  \includegraphics[width=0.95\linewidth]{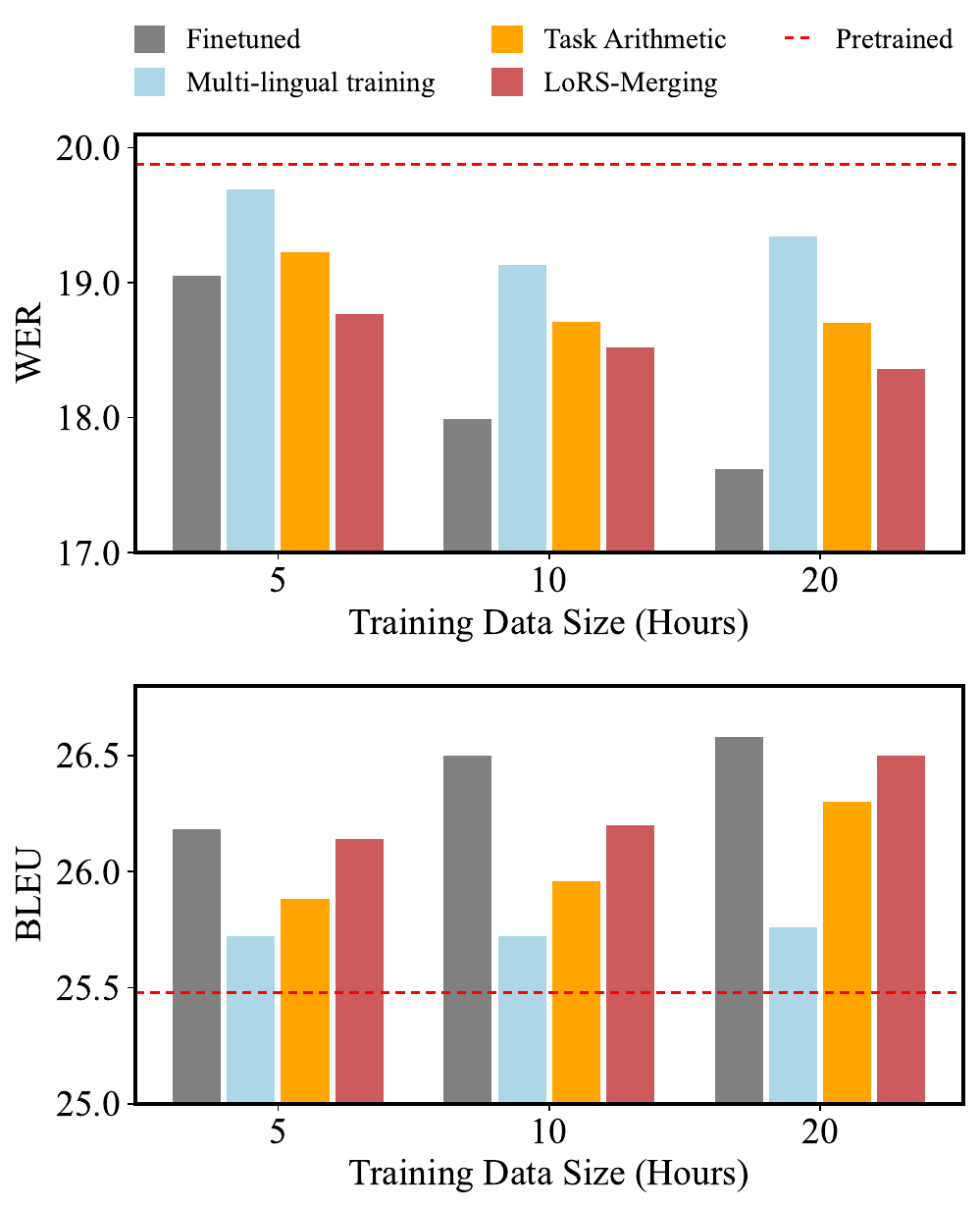}
  \caption{Performance variation against different training data sizes (number of hours for each language) on ASR (top) and ST (bottom) tasks.}
  \label{fig:data_scale}
  \vspace{-0.4cm}
\end{figure}

We then demonstrate the robustness of merging methods to different training data sizes for both tasks. 
Fig. \ref{fig:data_scale} shows the WER (top) and BLEU (bottom) scores for ASR and ST at different data scales, respectively.
As the data scale increases, the performance of multi-lingual training does not always improve.
This is because the multi-lingual capabilities of pretrained models are already near convergence, and only meticulous training can further improve performance. Increased training data amplifies both language interference and the negative effects of uniform training configurations, thereby offsetting the gains from increased data.
Furthermore, the performance loss of model merging increases with data scale, compared to finetuned models.
It can be explained by the fact that larger training data tends to increase the divergence in the optimisation trajectories of different finetuned models, resulting in the breakdown of linear mode connectivity, which leads to a greater performance loss.
Moreover, LoRS-Merging still achieves obvious and stable improvement compared to TA.


\subsection{Analysis of Model Redundancy}
\begin{figure}[t]
  \vspace{-0.25cm}
  \includegraphics[width=1.0\linewidth]{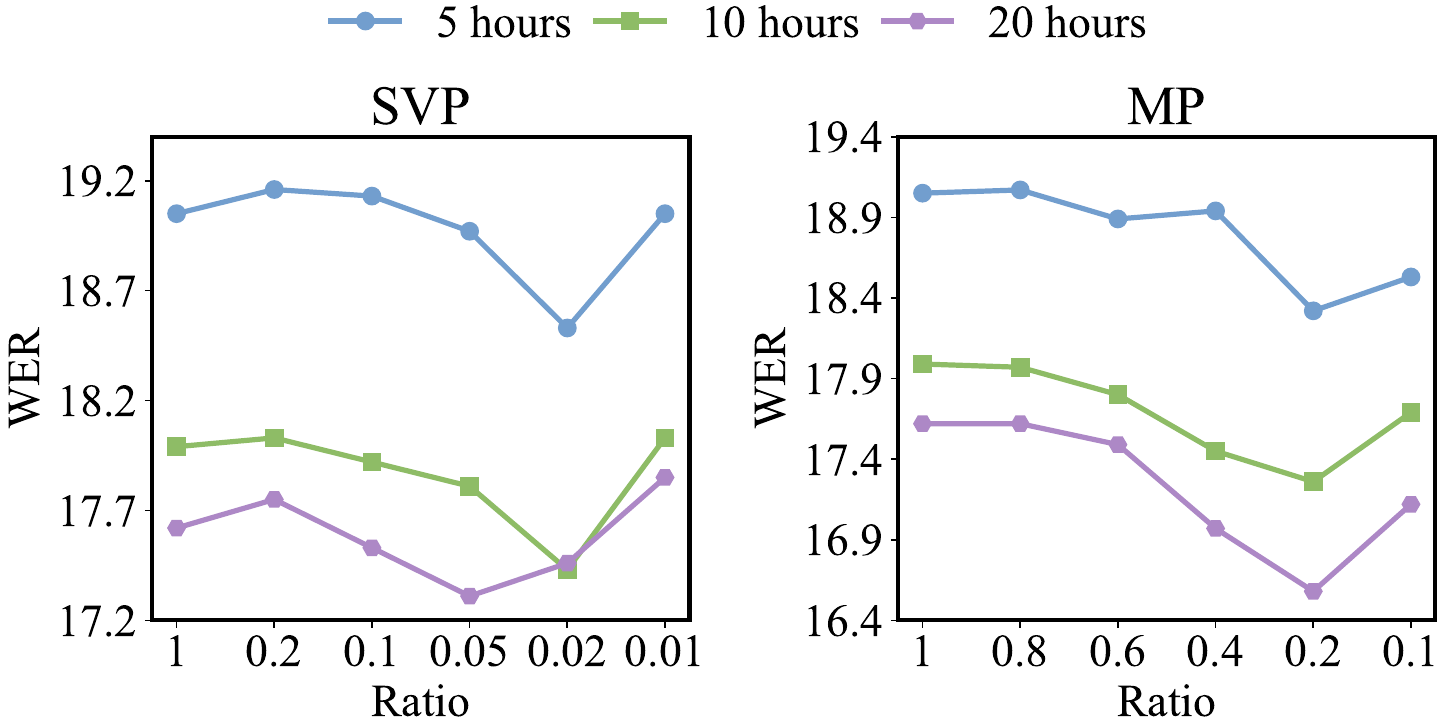}
  \caption{Model performance against the retain ratio in SVP (left) and MP (right) for ASR finetuned models. Three different training data sizes are used.}
  \label{fig:svp_mp_asr}
  \vspace{-0.4cm}
\end{figure}

Furthermore, we justify the necessity of SVP and MP to remove model redundancy by showing the model performance against the pruning ratio of finetuned models for ASR as shown in Fig. \ref{fig:svp_mp_asr}.
As shown, both SVP and MP significantly improve the performance of finetuned models, indicating the presence of substantial redundancy in the structure and details of the finetuned models, respectively. The model performance reaches the best at a high pruning level, indicating that the redundancy is particularly large for ASR. We observed a much smaller redundancy in ST, which also explains the observation that LoRS-Merging achieves more salient improvement on ASR than ST.
Moreover, redundancy increases with training data, possibly due to the accumulation of gradient noise during training.
MP achieves greater performance gains than SVP, indicating more redundancy at the detail level, which is better addressed by fine-grained MP.


\section{Conclusion}
This paper explores model merging for multi-lingual ASR and ST on pretrained speech models and proposes the LoRS-Merging approach. LoRS-Merging combines low-rank and sparse pruning to retain essential structures and reduce redundant parameters. Experiments across 10 languages show that LoRS-Merging effectively alleviates language interference and significantly outperforms multi-lingual multi-task training, sequential training, and other merging methods.



\section{Limitations}

There are three main limitations of this work. First, as a common limitation of all model merging methods, the same model structure is required across all tasks and languages. This is less of a concern under the current trend of using the same Transformer structure, but methods need to be developed in the future to accommodate subtle structural differences. Second, reasonably-sized training sets are required for each language, and low-resource languages may suffer from reduced improvements. Third, this work mainly explores the two most popular S2T tasks. Other possible tasks can be explored in future work, including spoken language understanding and speaker adaptation. 


\bibliography{custom}

\twocolumn[\newpage]

\appendix

\section{Details of the Experimental Setup}
\label{sec:appendix_setup}
For multi-lingual and multi-task training, a uniform training configuration is used across all languages and tasks.
For sequential training, considering that there are 5! = 120 possible sequences for 5 languages, and the optimal training configuration for the same language differs across sequences, the hyper-parameter search cost for sequential training is much higher than that for the model merging. To simplify the configuration, we select 5 sequences for the experiments, corresponding to all cyclic permutations of the language order ca-de-es-fr-it, and report the results from the sequence that yields the best average performance.

MP and SVP are applied to each linear layer.
The detailed hyper-parameter settings for each language are shown in Table \ref{tab:ASR-hyperparameter} for ASR and Table \ref{tab:ST-hyperparameter} for ST, respectively.

\begin{table}[ht]
  \caption{ASR hyper-parameters for high-resource languages.}
  \label{tab:ASR-hyperparameter}
  \centering
  \setlength{\tabcolsep}{2pt}
  \begin{adjustbox}{width=\columnwidth}
  \begin{tabular}{l|ccccc}
    \toprule
    \multicolumn{1}{l}{\multirow{2}{*}{\textbf{System}}} & \multicolumn{5}{c}{\textbf{ASR}} \\
     & ca & de & es & fr & it \\
    \midrule
    \textbf{Finetuned} \\
    learning rate & $1 \times 10^{-6}$ & $5 \times 10^{-8}$ & $1 \times 10^{-7}$ & $1 \times 10^{-6}$ & $5 \times 10^{-6}$ \\
    \midrule
    \textbf{Multi-lingual training} \\
    learning rate & \multicolumn{5}{c}{$1 \times 10^{-5}$} \\
    \midrule
    \textbf{Task Arithmetic} \\
    scaling factor $\lambda$ & \multicolumn{5}{c}{0.15} \\
    \midrule
    \textbf{LoRS-Merging} \\
    scaling factor $\lambda$ & \multicolumn{5}{c}{0.15} \\
    SVP ratio $r$ & 5\% & 3\% & 2\% & 1\% & 1\% \\
    MP ratio $p$ & 40\% & 60\% & 40\% & 10\% & 10\% \\
    \bottomrule
  \end{tabular}
  \end{adjustbox}
\end{table}

\begin{table}[ht]
  \caption{ST hyper-parameters for high-resource languages.}
  \label{tab:ST-hyperparameter}
  \centering
  \setlength{\tabcolsep}{2pt}
  \begin{adjustbox}{width=\columnwidth}
  \begin{tabular}{l|ccccc}
    \toprule
    \multicolumn{1}{l}{\multirow{2}{*}{\textbf{System}}} & \multicolumn{5}{c}{\textbf{ST}} \\
     & ca & de & es & fr & it \\
    \midrule
    \textbf{Finetuned} \\
    learning rate & $1 \times 10^{-6}$ & $2 \times 10^{-8}$ & $2 \times 10^{-8}$ & $5 \times 10^{-8}$ & $5 \times 10^{-8}$ \\
    \midrule
    \textbf{Multi-lingual training} \\
    learning rate & \multicolumn{5}{c}{$5 \times 10^{-9}$} \\
    \midrule
    \textbf{Task Arithmetic} \\
    scaling factor $\lambda$ & \multicolumn{5}{c}{0.15} \\
    \midrule
    \textbf{LoRS-Merging} \\
    scaling factor $\lambda$ & \multicolumn{5}{c}{0.15} \\
    SVP ratio $r$ & 5\% & 3\% & 5\% & 2\% & 1\% \\
    MP ratio $p$ & 60\% & 40\% & 20\% & 20\% & 20\% \\
    \bottomrule
  \end{tabular}
  \end{adjustbox}
\end{table}


\section{Results of Low-Resource Language Set}
\label{sec:appendix_low}

The results of the low-resource language set are shown in this section. Specifically, Table \ref{tab:multi-lingual-ASR-2} and \ref{tab:multi-lingual-ST-2} show the multi-lingual single-task training and merging for ASR and ST respectively.

\begin{table}[ht]
  \caption{Multi-lingual ASR model merging with the low-resource language set. Avg. denotes average WER.}
  \label{tab:multi-lingual-ASR-2}
  \centering
  \begin{adjustbox}{width=\columnwidth}
  \begin{tabular}{l|ccccc|c}
    \toprule
    \multicolumn{1}{l}{\multirow{2}{*}{\textbf{System}}} & \multicolumn{6}{c}{\textbf{WER$\downarrow$}} \\
     & id & nl & pt & ru & sv & Avg. \\
    \midrule
    Pretrained & 16.9 & 16.0 & 10.1 & 17.1 & 17.1 & 15.43 \\
    Finetuned & 15.0 & 14.8 & 9.7 & 16.8 & 14.7 & 14.20 \\
    \midrule
    Multi-lingual training & 16.7 & 15.5 & 10.0 & 17.0 & 16.6 & 15.14 \\
    \midrule
    Weight Averaging & 15.7 & 15.2 & 10.1 & 17.1 & 15.8 & 14.77 \\
    Task Arithmetic & 15.7 & 15.1 & 9.9 & 17.0 & 15.8 & 14.69 \\
    MP-Merging & 15.7 & 15.1 & 10.0 & 16.7 & 15.7 & 14.63 \\
    SVP-Merging & 15.7 & 15.1 & 9.9 & 16.9 & 15.7 & 14.65 \\
    LoRS-Merging & 15.7 & 15.1 & 9.7 & 16.8 & 15.6 & \textbf{14.57} \\
    \bottomrule
  \end{tabular}
  \end{adjustbox}
\end{table}

\begin{table}[ht]
  \caption{Multi-lingual ST model merging with the low-resource language set. Avg. denotes average BLEU.}
  \label{tab:multi-lingual-ST-2}
  \centering
  \begin{adjustbox}{width=\columnwidth}
  \begin{tabular}{l|ccccc|c}
    \toprule
    \multicolumn{1}{l}{\multirow{2}{*}{\textbf{System}}} & \multicolumn{6}{c}{\textbf{BLEU$\uparrow$}} \\
     & id & nl & pt & ru & sv & Avg. \\
    \midrule
    Pretrained & 32.5 & 31.6 & 43.3 & 35.5 & 32.1 & 35.00 \\
    Finetuned & 35.2 & 34.0 & 43.8 & 36.7 & 37.6 & 37.46 \\
    \midrule
    Multi-lingual training & 32.3 & 33.2 & 43.5 & 35.4 & 34.3 & 35.74 \\
    \midrule
    Weight Averaging & 33.6 & 32.2 & 43.2 & 35.3 & 34.2 & 35.70 \\
    Task Arithmetic & 33.9 & 32.8 & 43.1 & 35.5 & 34.3 & 35.92 \\
    MP-Merging & 33.8 & 32.8 & 43.5 & 35.8 & 34.0 & 35.98 \\
    SVP-Merging & 33.6 & 32.6 & 43.4 & 35.6 & 34.3 & 35.90 \\
    LoRS-Merging & 33.9 & 32.8 & 43.2 & 35.9 & 34.5 & \textbf{36.06} \\
    \bottomrule
  \end{tabular}
  \end{adjustbox}
\end{table}

Then, Table \ref{tab:multi-task-2} shows the multi-task single-language training and merging performance (c.f. compare to Table \ref{tab:multi-task} for high-resource languages).

\begin{table*}[ht]
  \caption{Multi-task model merging with the low-resource language set. WER/BLEU scores are averaged across languages.}
  \label{tab:multi-task-2}
  \centering
  \begin{adjustbox}{width=0.95\textwidth}
  \begin{tabular}{l|ccccc|c|ccccc|c}
    \toprule
    \multicolumn{1}{l}{\multirow{2}{*}{\textbf{System}}} & \multicolumn{6}{c}{\textbf{WER$\downarrow$}} & \multicolumn{6}{c}{\textbf{BLEU$\uparrow$}} \\
     & id & nl & pt & ru & sv & Avg. & id & nl & pt & ru & sv & Avg. \\
    \midrule
    Pretrained & 16.9 & 16.0 & 10.1 & 17.1 & 17.1 & 15.43 & 32.5 & 31.6 & 43.3 & 35.5 & 32.1 & 35.00 \\
    Finetuned & 15.0 & 14.8 & 9.7 & 16.8 & 14.7 & 14.20 & 35.2 & 34.0 & 43.8 & 36.7 & 37.6 & 37.46 \\
    \midrule
    Multi-task training & 15.4 & 15.0 & 9.3 & 16.6 & 14.3 & 14.12 & 35.3 & 33.7 & 43.6 & 36.2 & 35.8 & 36.92 \\
    \midrule
    Weight Averaging & 14.7 & 14.9 & 9.3 & 16.6 & 13.8 & 13.88 & 35.4 & 33.9 & 44.1 & 36.3 & 35.9 & 37.12 \\
    Task Arithmetic & 14.6 & 14.9 & 9.3 & 16.5 & 14.0 & 13.88 & 35.3 & 33.8 & 44.3 & 36.1 & 36.4 & 37.18 \\
    MP-Merging & 14.4 & 14.7 & 9.4 & 16.5 & 13.8 & 13.78 & 35.7 & 33.9 & 44.3 & 36.1 & 36.1 & 37.22 \\
    SVP-Merging & 14.6 & 14.8 & 9.2 & 16.4 & 13.9 & 13.80 & 35.3 & 33.9 & 44.3 & 36.2 & 36.3 & 37.20 \\
    LoRS-Merging & 14.4 & 14.7 & 9.2 & 16.4 & 13.8 & \textbf{13.72} & 35.6 & 33.9 & 44.3 & 36.3 & 36.4 & \textbf{37.30} \\
    \bottomrule
  \end{tabular}
  \end{adjustbox}
\end{table*}

Last, Table \ref{tab:multi-lingual multi-task-2} shows the results of multi-lingual and multi-task training and merging results for low-resource languages (compare to Table \ref{tab:multi-lingual multi-task} for high-resource languages.). LoRS-Merging achieved the best performance across all merging and training methods in all tables.

\begin{table*}[ht]
  \caption{Multi-lingual multi-task model merging with the low-resource language set. WER/BLEU scores are averaged across languages.}
  \label{tab:multi-lingual multi-task-2}
  \centering
  \begin{adjustbox}{width=0.95\textwidth}
  \begin{tabular}{l|ccccc|c|ccccc|c}
    \toprule
    \multicolumn{1}{l}{\multirow{2}{*}{\textbf{System}}} & \multicolumn{6}{c}{\textbf{WER$\downarrow$}} & \multicolumn{6}{c}{\textbf{BLEU$\uparrow$}} \\
     & id & nl & pt & ru & sv & Avg. & id & nl & pt & ru & sv & Avg. \\
    \midrule
    Pretrained & 16.9 & 16.0 & 10.1 & 17.1 & 17.1 & 15.43 & 32.5 & 31.6 & 43.3 & 35.5 & 32.1 & 35.00 \\
    Finetuned & 15.0 & 14.8 & 9.7 & 16.8 & 14.7 & 14.20 & 35.2 & 34.0 & 43.8 & 36.7 & 37.6 & 37.46 \\
    \midrule
    ML and MT training & 16.9 & 15.7 & 9.6 & 17.0 & 16.3 & 15.08 & 32.8 & 32.9 & 43.3 & 35.4 & 32.6 & 35.40 \\
    \midrule
    ML and MT Task Arithmetic & 16.4 & 15.5 & 9.6 & 16.8 & 15.7 & 14.79 & 33.7 & 33.1 & 43.2 & 35.7 & 34.9 & 36.12 \\
    ML and MT LoRS-Merging & 16.1 & 15.5 & 9.5 & 16.8 & 15.7 & 14.72 & 33.7 & 33.2 & 43.5 & 35.8 & 34.9 & \textbf{36.22} \\
    \midrule
    MT training & 15.4 & 15.0 & 9.3 & 16.6 & 14.3 & 14.12 & 35.3 & 33.7 & 43.6 & 36.2 & 35.8 & 36.92 \\
    $\hookrightarrow$ + ML Task Arithmetic & 16.0 & 15.5 & 9.5 & 16.9 & 15.4 & 14.66 & 34.1 & 32.8 & 43.7 & 35.6 & 33.3 & 35.90 \\
    $\hookrightarrow$ + ML LoRS-Merging & 16.1 & 15.3 & 9.4 & 16.8 & 15.3 & \textbf{14.57} & 34.2 & 32.7 & 43.8 & 35.8 & 33.5 & 36.00 \\
    \midrule
    ML training & 16.7 & 15.5 & 10.0 & 17.0 & 16.6 & 15.14 & 32.3 & 33.2 & 43.5 & 35.4 & 34.3 & 35.74 \\
    $\hookrightarrow$ + MT Task Arithmetic & 17.1 & 15.5 & 9.5 & 17.0 & 15.5 & 14.89 & 32.1 & 33.1 & 43.6 & 35.7 & 33.6 & 35.62 \\
    $\hookrightarrow$ + MT LoRS-Merging & 16.9 & 15.5 & 9.4 & 16.8 & 15.5 & 14.80 & 32.6 & 33.2 & 43.6 & 35.9 & 33.6 & 35.78 \\
    \bottomrule
  \end{tabular}
  \end{adjustbox}
\end{table*}


\section{Detailed Results on Multi-Task Model Merging}
\label{sec:appendix_detail}

Detailed per-language results of Table \ref{tab:multi-task} are shown in Table \ref{tab:multi-task-d}.

\begin{table*}[ht]
  \caption{Multi-task model merging with the high-resource language set. WER/BLEU scores are averaged across languages.}
  \label{tab:multi-task-d}
  \centering
  \begin{adjustbox}{width=0.95\textwidth}
  \begin{tabular}{l|ccccc|c|ccccc|c}
    \toprule
    \multicolumn{1}{l}{\multirow{2}{*}{\textbf{System}}} & \multicolumn{6}{c}{\textbf{WER$\downarrow$}} & \multicolumn{6}{c}{\textbf{BLEU$\uparrow$}} \\
     & ca & de & es & fr & it & Avg. & ca & de & es & fr & it & Avg. \\
    \midrule
    Pretrained & 20.6 & 19.6 & 14.7 & 24.5 & 19.4 & 19.88 & 21.1 & 24.1 & 28.6 & 26.8 & 26.8 & 25.48 \\
    Finetuned & 19.5 & 19.7 & 14.4 & 22.1 & 19.2 & 19.05 & 22.6 & 24.6 & 29.2 & 27.2 & 27.3 & 26.18 \\
    \midrule
    Multi-task training & 17.0 & 19.7 & 14.4 & 24.2 & 19.4 & 19.00 & 22.3 & 24.6 & 28.7 & 27.0 & 26.9 & 25.90 \\
    Sequential training & 16.7 & 19.4 & 14.3 & 24.6 & 19.4 & 18.95 & 22.9 & 24.8 & 28.7 & 27.3 & 26.9 & 26.12 \\
    \midrule
    Weight Averaging & 17.1 & 19.6 & 13.9 & 23.7 & 19.6 & 18.84 & 22.9 & 24.4 & 29.0 & 27.7 & 26.9 & 26.18 \\
    Task Arithmetic & 17.2 & 19.3 & 14.0 & 23.3 & 19.7 & 18.76 & 23.4 & 24.5 & 28.9 & 27.7 & 27.0 & 26.30 \\
    TIES-Merging & 17.7 & 19.5 & 14.4 & 23.6 & 17.4 & 18.60 & 23.1 & 24.5 & 29.1 & 27.7 & 27.5 & 26.38 \\
    DARE & 17.5 & 19.4 & 14.2 & 23.5 & 18.6 & 18.71 & 23.2 & 24.5 & 29.0 & 27.6 & 27.1 & 26.28 \\
    TSV-M & 17.8 & 19.4 & 14.3 & 23.7 & 17.9 & 18.70 & 23.0 & 24.7 & 29.2 & 27.7 & 27.4 & 26.40 \\
    LoRS-Merging & 17.3 & 19.4 & 14.1 & 23.1 & 17.7 & \textbf{18.39} & 23.3 & 24.6 & 29.3 & 28.0 & 27.6 & \textbf{26.56} \\
    \bottomrule
  \end{tabular}
  \end{adjustbox}
\end{table*}

\end{document}